# Rich Nature of Van Hove Singularities in Kagome Superconductor CsV$_3$Sb$_5$

Yong Hu[1,#,*], Xianxin Wu[2,#], Brenden R. Ortiz[3,#], Sailong Ju[1], Xinlong Han[4], J. Z. Ma[5], N. C. Plumb[1], Milan Radovic[1], Ronny Thomale[6,7], S. D. Wilson[3], Andreas P. Schnyder[2,*], and M. Shi[1,*]

[1]*Swiss Light Source, Paul Scherrer Institute, CH-5232 Villigen PSI, Switzerland*
[2]*Max-Planck-Institut für Festkörperforschung, Heisenbergstrasse 1, D-70569 Stuttgart, Germany*
[3]*Materials Department and California Nanosystems Institute, University of California Santa Barbara, Santa Barbara, California 93106, USA*
[4]*Department of Physics and Center of Theoretical and Computational Physics, University of Hong Kong, Hong Kong, China.*
[5]*Department of Physics, City University of Hong Kong, Kowloon, Hong Kong, China*
[6]*Institute for Theoretical Physics, University of Würzburg, Am Hubland, D-97074 Würzburg, Germany*
[7]*Department of Physics and Quantum Centers in Diamond and Emerging Materials (QuCenDiEM) group, Indian Institute of Technology Madras, Chennai 600036, India*

[#]These authors contributed equally to this work.

[*]To whom correspondence should be addressed:

Y.H. (yong.hu@psi.ch), A.P.S. (a.schnyder@fkf.mpg.de), M.S. (ming.shi@psi.ch).

**The recently discovered layered kagome metals AV$_3$Sb$_5$ (A=K, Rb, Cs) exhibit diverse correlated phenomena, which are intertwined with a topological electronic structure with multiple van Hove singularities (VHSs) in the vicinity of the Fermi level. As the VHSs with their large density of states enhance correlation effects, it is of crucial importance to determine their nature and properties. Here, we combine polarization-dependent angle-resolved photoemission spectroscopy with density functional theory to directly reveal the sublattice properties of *3d*-orbital VHSs in CsV$_3$Sb$_5$. Four VHSs are identified around the M point and three of them are close to the Fermi level, with two having sublattice-pure and one sublattice-mixed nature. Remarkably, the VHS just below the Fermi level displays an extremely flat dispersion along MK, establishing the experimental discovery of higher-order VHS. The characteristic intensity modulation of Dirac cones around K further demonstrates the sublattice interference embedded in the electronic structure. The crucial insights into the electronic structure, revealed by our work, provide a solid starting point for the understanding of the intriguing correlation phenomena in the kagome metals AV$_3$Sb$_5$.**

Transition-metal based kagome materials, hosting corner-sharing triangles, offer an exciting platform to explore intriguing correlated [1-3] and topological phenomena [4-6], including quantum spin liquid [7-11], unconventional superconductivity [12-15], Dirac/Weyl semimetals [16-18] and charge density wave (CDW) order [13-15,19,20]. Their emergence originates from the inherent features of the kagome lattice: substantial geometric spin frustration, flat bands, Dirac cones and van Hove singularities (VHSs) at different electron fillings. Recently, a new family of kagome metal $AV_3Sb_5$ (A= K, Rb, Cs) [21] with V kagome nets, was found to feature a $\mathbb{Z}_2$ topological band structure [22,23] and superconductivity was realized with a maximum $T_c$ of 2.5 $K$ at ambient pressure [19]. Moreover, they exhibit CDW order below $T_{CDW}$≈78-103 $K$ [24-27]. Aside from the translational symmetry breaking in this CDW phase, the breaking of additional symmetries, i.e., rotation and time-reversal, was observed upon cooling down towards $T_c$ [24,27,28]. Despite evidence supporting a nodeless gap from magnetic penetration depth measurements [29], double superconducting domes under pressure [30-32], a large residual in the thermal conductivity [33] and an edge supercurrent in $Nb/K_{1-x}V_3Sb_5$ suggest electronically driven and unconventional superconductivity [34]. It's widely believed that these exotic correlated phenomena are intimately connected with the multiple VHSs in the vicinity of the Fermi level [35-37].

The characteristics of VHS bands are crucial in determining the Fermi surface instabilities [12-15,38-42]. From the perspective of band dispersion around the saddle point, VHSs can be classified into two types: conventional and higher-order [43,44], as shown in Figs. 1e (i) and (ii). The higher-order VHS displays a flat dispersion along one direction with less pronounced Fermi surface nesting, generating a power-law divergent density of states (DOS) in two dimensions (2D) instead of a logarithmic divergent one [43,44]. Moreover, VHSs in kagome lattices possess distinct sublattice features: sublattice pure (*p*-type) and sublattice mixing (*m*-type), as shown in Fig. 1c. They induce an effective reduction of the local Coulomb interaction, thereby enhancing the role of non-local Coulomb terms [13,14,35]. Therefore, the nature of VHSs is pivotal to understand correlated phenomena, but still remains elusive in the kagome metals $AV_3Sb_5$ so far.

In this work, we perform a comprehensive study on the electronic structure of $CsV_3Sb_5$ by combining polarization-dependent angle-resolved photoemission spectroscopy (ARPES) measurements with Density Functional theory (DFT). The diverse nature of the four VHSs in the vicinity of the Fermi level ($E_F$) is directly revealed. We observe three VHSs around the M point below the $E_F$, formed by

Vanadium *3d* orbitals. Two of them are of conventional *p*-type, while the other one is of higher-order *p*-type. In addition, we find a conventional *m*-type VHS slightly above $E_F$ from our theoretical calculations. Furthermore, we show that the sublattice features are also embedded in the Dirac cone around the K point, exhibiting characteristic intensity modulations under various polarization conditions. Our study provides crucial insights into the electronic structure, thereby laying down the basis for a solid understanding of the correlation phenomena in the kagome metals $AV_3Sb_5$.

$CsV_3Sb_5$ crystalizes in a layered hexagonal lattice consisting of alternately stacked V-Sb sheets and Cs layers. Each V-Sb sheet contains a 2D vanadium kagome net interweaved by a hexagonal lattice of Sb atoms (Fig. 1a). The vanadium kagome lattice, shown in Fig. 1b, hosts three distinct sublattices located at 3f Wyckoff positions. The toy band of the kagome lattice displays two different types of VHSs: *p*-type and *m*-type. For the *p*-type VHS, the states near the three M points are contributed by mutually different sublattices, while the eigenstates of the *m*-type VHS are equally distributed over mutually different sets of two sublattices for each M point, as illustrated in Fig. 1c. The band structure of $CsV_3Sb_5$ from DFT calculations is displayed in Fig. 1d, where four VHS points occur at M in the vicinity of $E_F$ (indicated by the red arrows and labeled as VHS1-4). Interestingly, VHS1 exhibits a much flatter dispersion along MK compared with the orthogonal direction MΓ. Further analysis shows that the quadratic contribution along MK is substantially reduced, indicating that VHS1 realizes a higher-order VHS (see Supplementary Materials for details). This is in contrast to the other three VHSs near $E_F$ which are all of conventional type with dominant quadratic dispersions in both directions. Motived by these observations, we perform ARPES measurements to investigate the electronic structure and focus on the nature of the VHSs in $CsV_3Sb_5$ which we infer from the polarization analysis.

The overall band dispersion and constant energy contours of $CsV_3Sb_5$ obtained via ARPES experiment are summarized in Fig. 2. The evolution of the electronic bands with different binding energy in Fig. 2a display sophisticated structures, including Sb contributed electron pockets near the zone center and kagome-derived Dirac cones at the K points with binding energy ~0.27 *eV* (Figs. 2a and 2b). To reveal the VHS around $\bar{M}$ point, we display the band structure along the $\bar{\Gamma}$ - $\bar{K}$ - $\bar{M}$ - $\bar{\Gamma}$ direction in Figs. 2c and 2d, which are in good agreement with our DFT calculations (Fig. 2e). From the dispersion around $\bar{M}$ point, we clearly identify three saddle points, denoted by VHS1, VHS2 and VHS3. Particularly, VHS1 and VHS2 are just slightly below $E_F$ with strong intensity. Remarkably, VHS1 exhibits a pronounced flat dispersion that extends over more than half of the $\bar{M}$ - $\bar{K}$ path. Indeed, by fitting the experimental spectra we find that the quadratic term is substantially smaller than the quartic one, revealing the higher-order nature of VHS1 (see Supplementary Fig. S1). Notably, the measured dispersion around VHS1 is much flatter than the theoretical one, indicating that renormalizations due to interactions enhance the higher-order nature of VHS1.

After directly identifying the VHSs in CsV$_3$Sb$_5$ from the band dispersion, we now turn to determining the sublattice nature of VHSs from the orbital symmetries by employing polarization-dependent ARPES measurements (see Supplementary Materials for experimental details). According to the selection rules in photoemission, bands can be selectively detected depending on their symmetry with respect to given mirror planes of the geometry [45]. Specifically, even- (resp. odd-) parity orbitals with respect to a mirror plane will be detected by the polarization whose electric field vector is in (resp. out of) the mirror plane. Our ARPES geometry is sketched in Fig. 3a. Polarization-dependent measurements were performed on the band structures along two orthogonal paths, $\bar{\Gamma}$ - $\bar{K}$ - $\bar{M}$ - $\bar{K}$ and $\bar{\Gamma}$ - $\bar{M}$ - $\bar{\Gamma}$ directions (Fig. 3b). We first employ circular polarization that can detect both even- and odd-parity orbitals to map out the full dispersions, as shown in Figs. 3c and 3d. To determine the orbital character of the bands, we further adopt linear horizontal (LH) polarization (Figs. 3e and 3f) and linear vertical (LV) (Figs. 3g and 3h) polarization. The electric field vector of LH and LV polarized light is in and out the mirror plane, respectively. Thus, when aligning the $\bar{G}$ - $\bar{M}$ direction of the sample to along the analyzer slit, $d_{xz}$, $d_{z^2}$ and $d_{x^2-y^2}$ are all of even symmetry with respect to the mirror plane and are detectable in the LH geometry. However, $d_{yz}$ and $d_{xy}$ are odd with respect to the mirror plane, and thus photoemission signals are only detectable in the LV geometry. Likewise, when the slit is along the $\bar{G}$ - $\bar{K}$ direction, the orbitals can be selected accordingly (see supplementary Table I). Based on these selection rules (see supplementary Fig. S2 for details of the matrix element analysis), the orbital characters of the bands constituting the VHSs below $E_F$ can be clearly identified, as shown in Fig. 3i. The VHS1, VHS2 and VHS3 are attributed to $d_{x^2-y^2}/d_{z^2}$, $d_{yz}$ an $d_{xy}$ orbitals, respectively.

With the experimental determination of the orbital character around the $\bar{M}$ point, we plot the theoretical orbital-resolved band dispersion in Fig. 3j originating from sublattice A (Fig. 1b), which is invariant under mirror reflection $M_{xz}$ and $M_{yz}$. Comparing the orbital characters around M point in

Figs. 3i and 3j, we find a good agreement between the experimental and theoretical results. Furthermore, the states at VHS1, VHS2 and VHS3 at $\bar{M}$ point, characterized by $A_g$, $B_{2g}$ and $B_{1g}$ irrep., are solely attributed to the A sublattice and inversion-even, confirming their *p*-type nature (Fig.3k). The band top of $d_{xz}$ band at $\bar{M}$ is above $E_F$ and beyond experimental observation (Figs. 3i and 3j). However, our calculations show that this band belongs to $B_{1u}$ irrep. (inversion-odd, Fig. 3k) and is attributed to a mixed contribution from sublattices B and C, implying its *m*-type nature. The four *p*-type and *m*-type VHSs, especially, the three of them (the VHS1, VHS2 and VHS4) are close to the Fermi level, suggesting that these VHSs with their large DOS and nontrivial sublattice and higher-order natures play a key role in driving the exotic correlated electronic states in $AV_3Sb_5$.

We further discuss the sublattice features of the Dirac cone bands around the $\bar{K}$ point. The polarization dependent intensity patterns of the Dirac cones can convey the phase information of electronic wave functions, providing a way to determine the chirality of the Dirac cone [6,46,47]. Figure 4a displays representative constant energy contours of $CsV_3Sb_5$ around the $\bar{K}$ point. The spectral intensity is strongly modulated around the kagome-derived Dirac cone for the LH polarization (Fig. 4b), with the maximum and minimum along the $\bar{\Gamma}$ - $\bar{K}$ direction but at opposite momentum direction above [Fig. 4a(i)] and below [Fig. 4a(iii)] the Dirac point [Figs. 4a (ii) and 4b]. Similar behavior is observed in LV polarization but in a reversed fashion (Figs. 4c and 4d). The intensity modulation around the Dirac cone, mimicking the case of graphene [47] and the kagome metal FeSn [6], indicates the chirality of the kagome-derived Dirac fermions in $CsV_3Sb_5$. The spectral intensity patterns (Figs. 4a and 4c) can be excellently reproduced in a spectral simulation based on sublattice interference of kagome initial states (Fig. 4e), further illustrating the sublattice interference embedded in the band structure of $CsV_3Sb_5$.

Our ARPES measurements, combined with DFT calculations, reveal the different natures of the four *3d*-orbital VHSs near $E_F$ and the chiralities of the kagome-derived Dirac cones in $CsV_3Sb_5$. These are general features in the family of kagome metals $AV_3Sb_5$ and have important physical implications. For example, the nontrivial sublattice texture of the *p*-type VHSs leads, via sublattice interference, to a suppression of local Hubbard interactions and promotes the relevance of non-local Coulomb terms [13,35]. The bands from the conventional *p*-type VHS2 feature a good Fermi surface nesting, which could be responsible for the observed CDW order. The higher-order *p*-type VHS1, on the other hand, exhibits less pronounced Fermi surface nesting with large DOS but can promote a nematic order [43,44], which may provide a possible explanation for the additional crystal symmetry breaking in the CDW phase at lower temperature. The appearance of multiple types of VHSs near $E_F$ including both *p*-type and *m*-type, derived from the multi-orbital nature, can induce a rich competition for various pairing instabilities and thus generate numerous different orders depending on small changes in the electron filling [35-37,43,44]. The kagome metals $AV_3Sb_5$ offer the tantalizing opportunity to access and tune these orders via carrier doping or external pressure [30,31,48-52], which remains to be further investigated both experimentally and theoretically.

**References:**

[1] M. R. Norman, Colloquium: Herbertsmithite and the search for the quantum spin liquid, *Rev. Mod. Phys.* **88**, 041002 (2016).

[2] A Mielke, Ferromagnetic ground states for the Hubbard model on line graphs, J. Phys. A: Math. Gen. 24, L73–L77 (1991).

[3] F. Pollmann, P. Fulde, and K. Shtengel, Kinetic Ferro-magnetism on a Kagome Lattice, *Phys. Rev. Lett.* **100**, 136404 (2008).

[4] L. Ye, M. Kang, J. Liu, F. Cube, C. R. Wicker, T. Suzuki, C. Jozwiak, A. Bostwick, E. Rotenberg, D. C. Bell, L. Fu, R. comin, and Joseph G. checkelsky, Massive Dirac fermions in a ferromagnetic kagome metal, *Nature* (London) **555**, 638–642 (2018).

[5] J.-X. Yin, W. Ma, T. A. Cochran, X. Xu, S. S. Zhang, H.-J. Tien, N. Shumiya, G. Cheng, K. Jiang, B. Lian, Z. Song, G. Chang, I. Belopolski, D. Multer, M. Litskevich, Z.-J. Cheng, X. P. Yang, B. Swidler, H. Zhou, H. Lin, T. Neupert, Z. Wang, N. Yao, T.-R. Chang, S. Jia, and M. Z. Hasan, Quantum-limit Chern topological magnetism in $TbMn_6Sn_6$, *Nature* (London) **583**, 533–536 (2020).

[6] M. Kang, L. Ye, S. Fang, J.-S. You, A. Levitan, M. Han, J. I. Facio, C. Jozwiak, A. Bostwick, E. Rotenberg, M. K. Chan, R. D. McDonald, D. Graf, K. Kaznatcheev, E. Vescovo, D. C. Bell, E. Kaxiras, J. Brink, M. Richter, M. P. Ghimire, J. G. Checkelsky, and R. Comin, Dirac fermions and flat bands in the ideal kagome metal FeSn, *Nat. Mater.* **19**, 163–169 (2020).

[7] D. Wulferding, P. Lemmens, P. Scheib, J. Roder, P. Mendels, S. Chu, T. Han, and Y. S. Lee, Interplay of thermal and quantum spin fluctuations in the kagome lattice compound herbertsmithite, *Phys. Rev. B* **82**, 144412 (2010).

[8] S. Yan, D. A. Huse, and S. R. White, Spin-Liquid Ground State of the S = 1/2 Kagome Heisenberg Antiferromagnet, *Science* **332**, 1173–1176 (2011).

[9] T.-H. Han, J. S. Helton, S. Chu, D. G. Nocera, J. A. Rodriguez- Rivera, C. Broholm, and Y. S. Lee, Fractionalized excitations in the spin-liquid state of a kagome-lattice antiferromagnet, *Nature* (London) **492**, 406–410 (2012).

[10] T. Han, S. Chu, and Y. S. Lee, Refining the Spin Hamiltonian in the Spin-1/2 Kagome Lattice Antiferromagnet $ZnCu_3(OH)_6Cl_2$ Using Single Crystals, *Phys. Rev. Lett.* **108**, 157202 (2012).

[11] M. Fu, T. Imai, T.-H. Han, and Y. S. Lee, Evidence for a gapped spin-liquid ground state in a kagome Heisenberg antiferromagnet, *Science* **350**, 655–658 (2015).

[12] W. -H. Ko, P. A. Lee, and X. -G. Wen, Doped kagome system as exotic superconductor, *Phys. Rev. B* **79**, 214502 (2009).

[13] M. L. Kiesel, and R.Thomale, Sublattice interference in the kagome Hubbard model, *Phys. Rev. B* **86**, 121105(R) (2012).

[14] M. L. Kiesel, C. Platt, and R. Thomale, Unconventional Fermi Surface Instabilities in the Kagome Hubbard Model, *Phys. Rev. Lett.* **110**, 126405 (2013).

[15] W.-S. Wang, Z.-Z. Li, Y.-Y. Xiang, and Q.-H. Wang, Competing electronic orders on kagome lattices at van Hove filling, *Phys. Rev. B* **87**, 115135 (2013).

[16] E. Liu et al., Giant anomalous Hall effect in a ferromagnetic kagome-lattice semimetal, *Nat. Phys.* **14**, 1125 (2018).

[17] N. Morali, R. Batabyal, P. K. Nag, E. Liu, Q. Xu, Y. Sun, B. Yan, C. Felser, N. Avraham, and H. Beidenkopf, Fermi-arc diversity on surface terminations of the magnetic Weyl semimetal $Co_3Sn_2S_2$, *Science* **365**, 1286–1291 (2019).

[18] D.-F. Liu, A.-J. Liang, E.-K. Liu, Q.-N. Xu, Y.-W. Li, C. Chen, D. Pei, W.-J. Shi, S. K. Mo, P. Dudin, T. Kim, C. Cacho, G. Li, Y. Sun, L. X. Yang, Z. K. Liu, S. P. Parkin, C. Felser, Y. L. Chen, Magnetic Weyl semimetal phase in a Kagomé crystal, *Science* **365**, 1282–1285 (2019).

[19] H.-M. Guo, and M. Franz, Topological insulator on the kagome lattice, *Phys. Rev. B* **80**, 113102 (2009).

[20] S. Isakov, S. Wessel, R. Melko, K. Sengupta, and Y. B. Kim, Hard-Core Bosons on the Kagome Lattice: Valence-Bond Solids and Their Quantum Melting, *Phys. Rev. Lett.* **97**, 147202 (2006).

[21] B. R. Ortiz, L. C. Gomes, J. R. Morey, M. Winiarski, M. Bordelon, J. S. Mangum, I. W. Oswald, J. A. Rodriguez-Rivera, J. R. Neilson, S. D. Wilson, New kagome prototype materials: discovery of $KV_3Sb_5$, $RbV_3Sb_5$, and $CsV_3Sb_5$, *Phys. Rev. Mater.* **3**, 094407 (2019).

[22] B. R. Ortiz, S. M. Teicher, Y. Hu, J. L. Zuo, P. M. Sarte, E. C. Schueller, A. M. Abeykoon, M. J. Krogstad, S. Rosenkranz, R. Osborn, R. Seshadri, L. Balents, J. He, and S. D. Wilson, $CsV_3Sb_5$: A $\mathbb{Z}_2$ Topological Kagome Metal with a Superconducting Ground State, *Phys. Rev. Lett.* **125**, 247002 (2020).

[23] Y. Hu, S. M. L. Teicher, B. R. Ortiz, Y. Luo, S. Peng, L. Huai, J. Z. Ma, N. C. Plumb, S. D. Wilson, J.-F. He, and M. Shi, Charge-order-assisted topological surface states and flat bands in the kagome superconductor $CsV_3Sb_5$, Preprint at arXiv:2104.12725 (2021).

[24] Y.-X. Jiang, J.-X. Yin, M. M. Denner, N. Shumiya, B. R. Ortiz, J. He, X.-X. Liu, S.-T. S. Zhang, G.-Q. Chang, I. Belopolski, Q. Zhang, T. A. Cochran, D. Multer, M. Litskevich, Z.-J. Cheng, X. P. Yang, Z. Wang, R. Thomale, T. Neupert, S. D. Wilson, and M. Zahid Hasan, Discovery of unconventional chiral charge order in kagome superconductor $KV_3Sb_5$, Preprint at arXiv:2012.15709 (2020).

[25] H.-X. Tan, Y.-Z. Liu, Z.-Q. Wang, and B.-H. Yan, Charge density waves and electronic properties of superconducting kagome metals, Preprint at arXiv:2103.06325 (2021).

[26] Z. Liang, X. Hou, F. Zhang, W. Ma, P. Wu, Z. Zhang, F. Yu, J. -J. Ying, K. Jiang, L. Shan, Z. Wang, X. -H. Chen, Three-dimensional charge density wave and robust zero-bias conductance peak inside the superconducting vortex core of a kagome superconductor $CsV_3Sb_5$, Preprint at arXiv:2103.04760 (2021).

[27] H. Chen, B. Hu, Z. Zhao, J. Yuan, Y. Xing, G. Qian, Z. Huang, G. Li1, Y. Ye, Q. Yin, C. Gong, Z. Tu, H. Lei, S. Ma, H. Zhang, S. Ni, H. Tan, C. Shen, X. Dong, B. Yan, Z. Wang, and H. -J. Gao, Roton pair density wave and unconventional strong-coupling superconductivity in a topological kagome metal, Preprint at arXiv:2103.09188 (2021).

[28] H. Li, H. Zhao, B. R. Ortiz, T. Park, M. Ye, L. Balents, Z. Wang, S. D. Wilson, and I. Zeljkovic, Rotation symmetry breaking in the normal state of a kagome superconductor $KV_3Sb_5$, Preprint at arXiv:2104.08209 (2021).

[29] W. Duan, Z. Nie, S. Luo, F. -H. Yu, B. R. Ortiz, L. Yin, H. Su, F. Du, A. Wang, Y. Chen, X. Lu, J. Ying, S. D. Wilson, X. Chen, Y. Song, and H. Yuan, Nodeless superconductivity in the kagome metal $CsV_3Sb_5$, Preprint at arXiv:2103.11796 (2021).

[30] K. Y. Chen, N. N. Wang, Q. W. Yin, Z. J. Tu, C. S. Gong, J. P. Sun, H. C. Lei, Y. Uwatoko, and J.-G. Cheng, Double superconducting dome and triple enhancement of $T_c$ in the kagome superconductor $CsV_3Sb_5$ under high pressure, Preprint at arXiv:2102.09328 (2021).

[31] Z. Zhang, Z. Chen, Y. Zhou, Y. Yuan, S. Wang, L. Zhang, X. Zhu, Y. Zhou, X. Chen, J. Zhou, and Z. Yang, Pressure-induced Reemergence of Superconductivity in Topological Kagome Metal $CsV_3Sb_5$, Preprint at arXiv:2103.12507 (2021).

[32] X. Chen, X. Zhan, X. Wang, J. Deng, X. -B. Liu, X. Chen, J. -G. Guo, and X. Chen, Highly-robust reentrant superconductivity in $CsV_3Sb_5$ under pressure, Preprint at arXiv:2103.13759 (2021).

[33] C. C. Zhao, L. S. Wang, W. Xia, Q. W. Yin, J. M. Ni, Y. Y. Huang, C. P. Tu, Z. C. Tao, Z. J. Tu, C. S. Gong, H. C. Lei, Y. F. Guo, X. F. Yang, and S. Y. Li, Nodal superconductivity and superconducting domes in the topological Kagome metal $CsV_3Sb_5$, Preprint at arXiv:2102.08356 (2021).

[34] Y. Wang, S.-Y. Yang, P. K. Sivakumar, B. R. Ortiz, S. M. L. Teicher, H. Wu, A. K. Srivastava, C. Garg, D. Liu, S. S. P. Parkin, E. S. Toberer, T. McQueen, S. D. Wilson, M. N. Ali, Proximity-induced spin-triplet superconductivity and edge supercurrent in the topological Kagome metal, $K_{1-x}V_3Sb_5$, Preprint at arXiv:2012.05898 (2020).

[35] X. Wu, T. Schwemmer, T. Müller, A. Consiglio, G. Sangiovanni, D. D. Sante, Y. Iqbal, W. Hanke, A. P. Schnyder, M. M. Denner, M. H. Fischer, T. Neupert, and R. Thomale, Nature of unconventional pairing in the kagome superconductors $AV_3Sb_5$, Preprint at arXiv:2104.05671 (2021).

[36] Y.-P. Lin, and R. M. Nandkishore, Complex charge density waves at Van Hove singularity on hexagonal lattices: Haldane-model phase diagram and potential realization in kagome metals $AV_3Sb_5$, Preprint at arXiv:2104.02725 (2021).

[37] T. Park, M. Ye, and L. Balents, Electronic instabilities of kagomé metals: saddle points and Landau theory, Preprint at arXiv:2104.08425 (2021).

[38] W. Kohn, and J. M. Luttinger, New mechanism for superconductivity, *Phys. Rev. Lett.* **15**, 524–526 (1965).

[39] T. M. Rice, and G. K. Scott, New mechanism for a charge-density-wave instability, *Phys. Rev. Lett.* **35**, 120–123 (1975).

[40] M. Fleck, A. M. Oleś, and L. Hedin, Magnetic phases near the Van Hove singularity in s- and d-band Hubbard models, *Phys. Rev. B* **56**, 3159–3166 (1997).

[41] J. K. González, Luttinger superconductivity in graphene, *Phys. Rev. B* **78**, 205431 (2008).

[42] K. Gofron, J. C. Campuzano, A. A. Abrikosov, M. Lindroos, A. Bansil, H. Ding, D. Koelling, and B. Dabrowski, Observation of an "Extended" Van Hove Singularity in $YBa_2Cu_4O_8$ by Ultrahigh Energy Resolution Angle-Resolved Photoemission, *Phys. Rev. Lett.* **73**, 3302 (1994).

[43] N. F.Q. Yuan, H. Isobe, and L. Fu, Magic of high-order van Hove singularity, *Nat. Commun.* **10**, 5769 (2019).

[44] L. Classen, A. V. Chubukov, C. Honerkamp, and M. M. Scherer, Competing orders at higher-order Van Hove points, *Phys. Rev. B* **102**, 125141 (2020).

[45] A. Damascelli, Z. Hussain, and Z.-X. Shen, Angle-resolved photoemission studies of the cuprate superconductors, *Rev. Mod. Phys.* **75**, 473-541 (2003).

[46] C. Hwang, C.-H. Park, D. A. Siegel, A. V. Fedorov, Steven G. Louie, and A. Lanzara, Direct measurement of quantum phases in graphene via photoemission spectroscopy, *Phys. Rev. B* **84**, 125422 (2011).

[47] Y. Liu, G. Bian, T. Miller, and T.-C. Chiang, Visualizing Electronic Chirality and Berry Phases in Graphene Systems Using Photoemission with Circularly Polarized Light, *Phys. Rev. Lett.* **107**, 166803 (2011).

[48] Y. Song, T. Ying, X. Chen, X. Han, Y. Huang, X. Wu, A. P. Schnyder, J.-G. Guo, and X. Chen, Enhancement of superconductivity in hole-doped CsV$_3$Sb$_5$ thin flakes, Preprint at arXiv:2105.09898 (2021).

[49] J. L. McChesney, A. Bostwick, T. Ohta, T. Seyller, K. Horn, J. González, and E. Rotenberg, Extended van Hove Singularity and Superconducting Instability in Doped Graphene, *Phys. Rev. Lett.* **104**, 136803 (2010).

[50] G. Li, A. Luican, J. M. B. L. Santos, A. H. C. Neto, A. Reina, J. Kong, and E. Y. Andrei, Observation of Van Hove singularities in twisted graphene layers, *Nat. Phys.* **6**, 109–113 (2010).

[51] P. Rosenzweig, H. Karakachian, D. Marchenko, K. Küster, and Ulrich Starke, Overdoping Graphene beyond the van Hove Singularity, *Phys. Rev. Lett.* **125**, 176403 (2020).

[52] A. J. H. Jones, R. Muzzio, P. Majchrzak, S. Pakdel, D. Curcio, K. Volckaert, D. Biswas, J. Gobbo, S. Singh, J. T. Robinson, K. Watanabe, T. Taniguchi, T. K. Kim, C. Cacho, N. Lanata, J. A. Miwa, P. Hofmann, J. Katoch, and S. Ulstrup, Observation of Electrically Tunable van Hove Singularities in Twisted Bilayer Graphene from NanoARPES, *Adv. Mater.* **32**, 2001656 (2020).

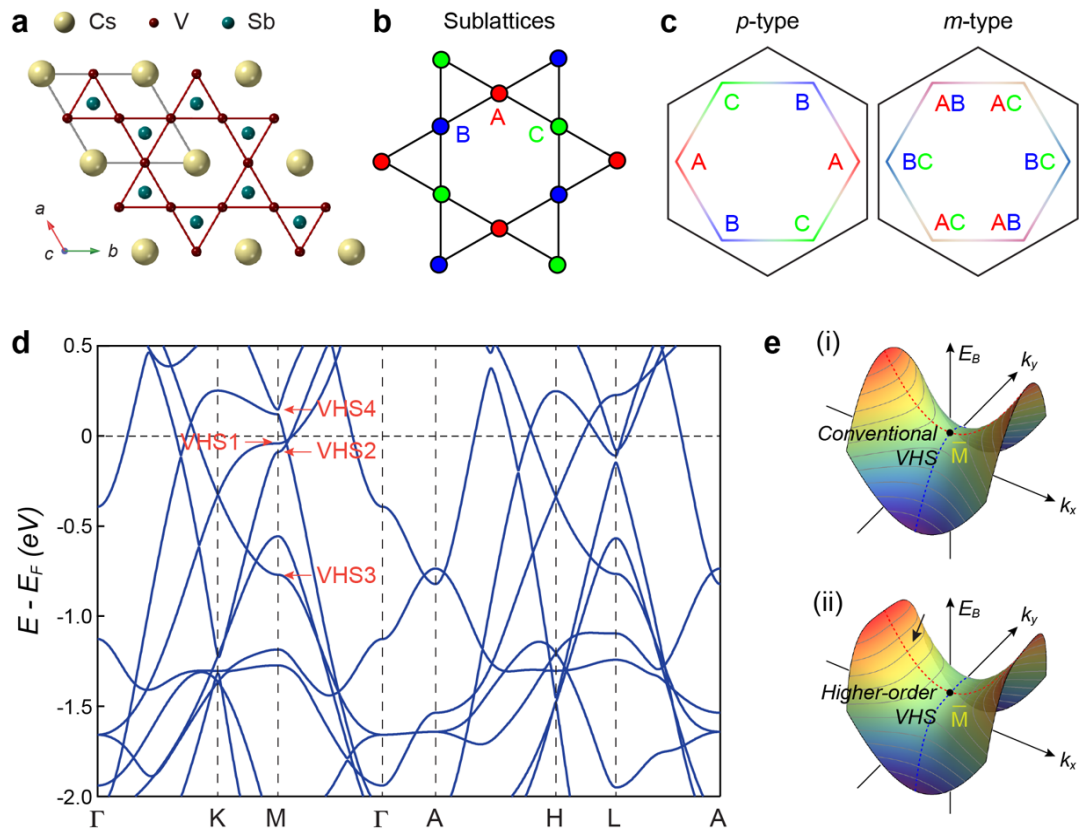

**Fig. 1 | Crystal structure, Kagome sublattices and van Hove singularities in kagome superconductors CsV$_3$Sb$_5$.**
**a** The Lattice structure of kagome metals CsV$_3$Sb$_5$ from top view. **b** Real space structure of the kagome vanadium planes. The red, blue and green coloring indicates the three kagome sublattices. **c** Two distinct types of sublattice decorated van Hove singularities (VHS) in CsV$_3$Sb$_5$, labeled as *p*-type (sublattice pure, left panel) and *m*-type (sublattice mixing, right panel). **d** Density functional theory calculated electronic structure of CsV$_3$Sb$_5$. The red arrows mark the VHSs. **e** Schematics of the conventional VHS (i) and higher-order VHS (ii) in two-dimensional electron systems. The grey curves in (e) indicate the constant energy contours that show markedly flat features along the $k_y$ direction in higher-order VHS, as highlighted by the black arrow.

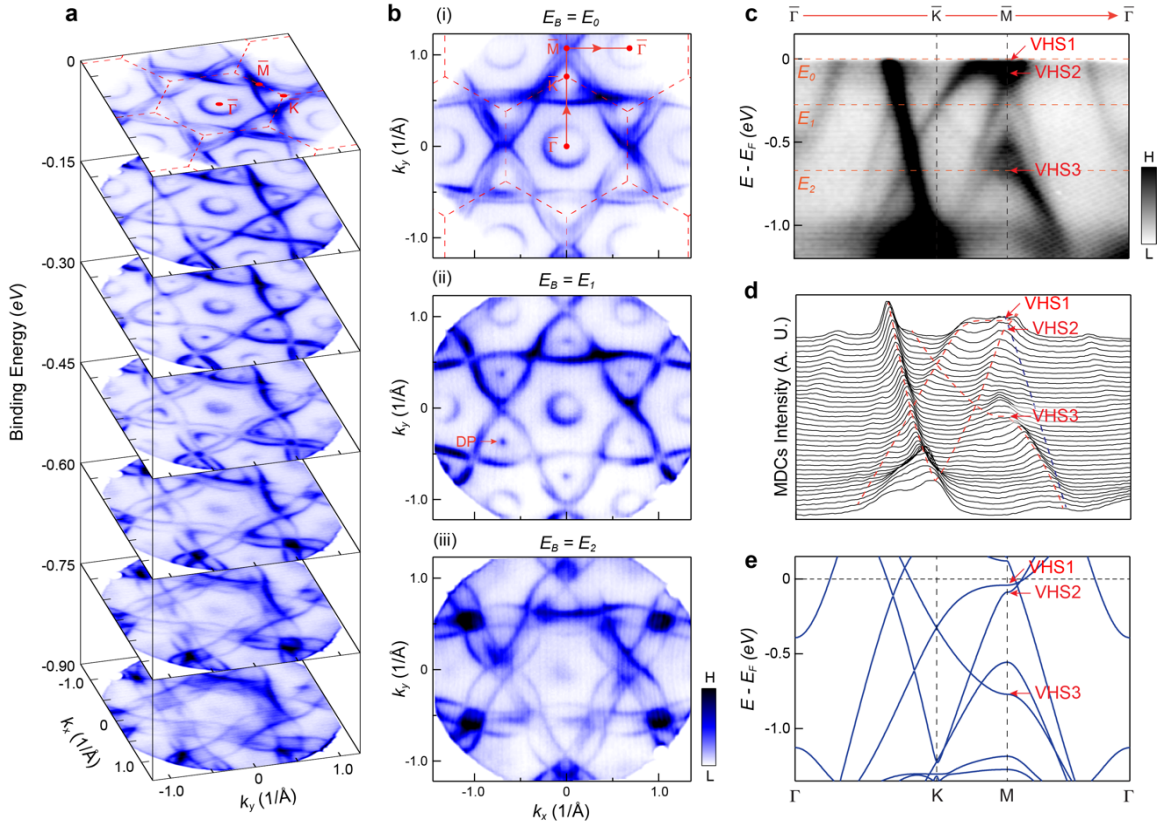

**Fig. 2 | Identification of multiple van Hove singularities in CsV$_3$Sb$_5$. a** Stacking plots of constant energy contours at different binding energies ($E_B$) show sophisticated band structure evolution as a function of energy. **b** Fermi surface (i), constant energy contours (CEC) at $E_B$ of the Dirac points (DP) (ii), and CEC at the VHS3 (iii). **c** Experimental band dispersion along the $\bar{\Gamma}$ - $\bar{K}$ - $\bar{M}$ - $\bar{\Gamma}$ direction. The momentum direction is indicted by the red arrows in [b (i)]. The orange dashed lines indicate the energy position of the Fermi level, Dirac cone and VHS3. **d** Momentum distribution curves (MDCs) of the high symmetry cut shown in (c). The dashed curves are guides to eye for band dispersions. **e** Calculated bands along the $\bar{\Gamma}$ - $\bar{K}$ - $\bar{M}$ - $\bar{\Gamma}$ direction. The arrows with different colors in (c-e) mark the multiple VHSs. All measurements were performed at 20 K (see Supplementary Fig. S4).

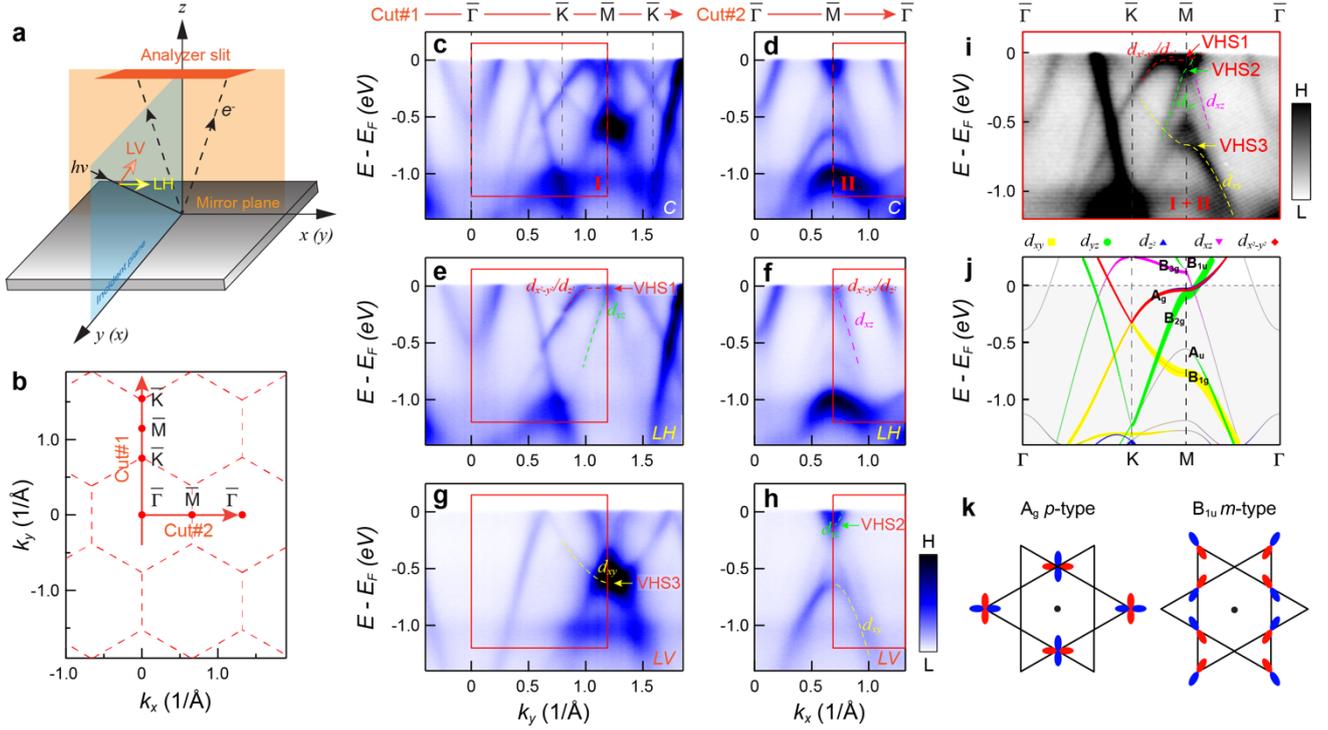

**Fig. 3| Determination of the orbital nature the kagome bands in CsV$_3$Sb$_5$. a** Experimental geometry of our polarization-dependent ARPES. **b** The two-dimensional projection of the Brillouin zone and the high-symmetry directions. **c,d** Band dispersions along the $\bar{\Gamma}$ - $\bar{K}$ - $\bar{M}$ - $\bar{K}$ [Cut#1, (c)] and $\bar{\Gamma}$ - $\bar{M}$ - $\bar{\Gamma}$ [Cut#2, (d)] directions, respectively. The momentum directions of the cuts are indicated by the red arrows in (b). The bands are measured with circularly polarized light, at 20 K. **e,f** and **g,h** Same as (c),(d), but probed with linear horizontal (LH) (e,f) and linear vertical (LV) (g,h) polarizations, respectively. **i** Experimental band structure along the $\bar{\Gamma}$ - $\bar{K}$ - $\bar{M}$ - $\bar{\Gamma}$ direction, with orbital characters marked. The momentum range is equal to the sum of the region I in (c) and region II in (d) selected by the red boxes. The dispersion is the same as the cut shown in Fig. 2(c). **j** Orbital character resolved band structure from calculations, with irreducible band representations labeled. **k** The sign structure (blue/red) and spatial orientation of the $d_{x^2-y^2}$-orbital A$_g$ *p*-type (inversion-even) and $d_{xz}$-orbital B$_{1u}$ *m*-type type (inversion-odd) VHS. The phase orbitals are plotted in the positive $k_z$ plane. The black dots in the center of the hexagons indicate the inversion centers.

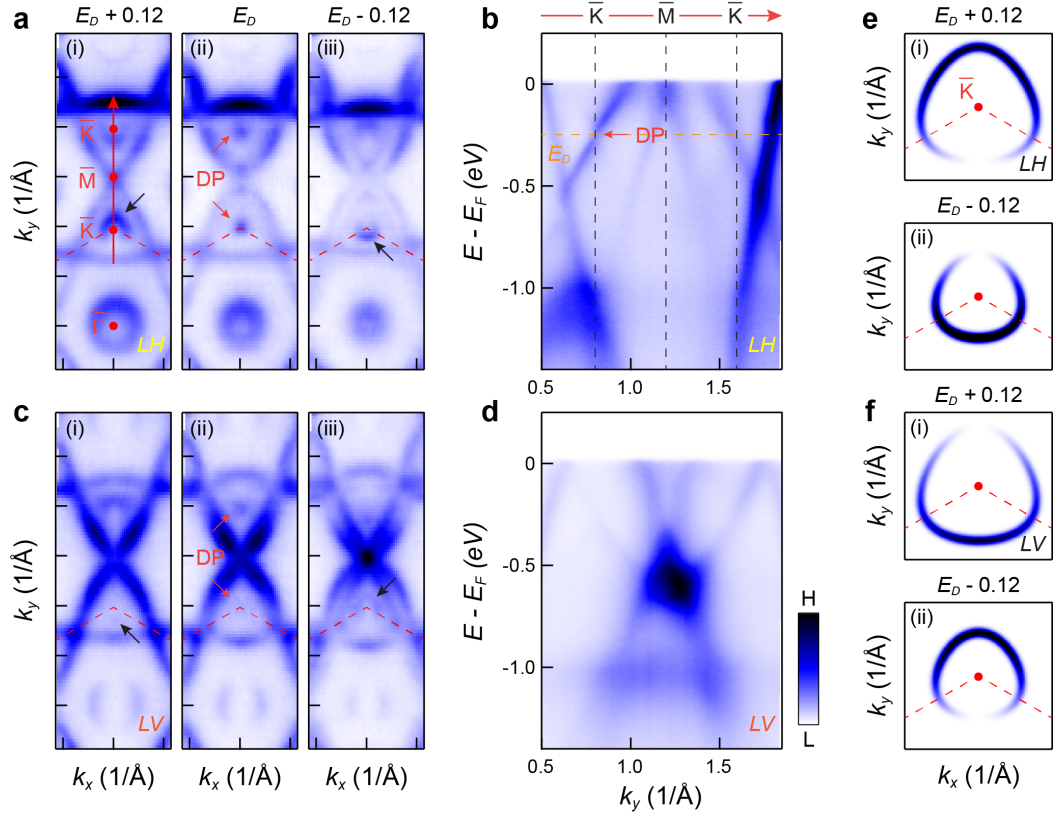

**Fig. 4 | Modulation of the photoemission intensity around the Dirac point in CsV$_3$Sb$_5$. a** Constant energy map above (i), at (ii) and below (iii) the Dirac cone, respectively. The black arrows indicate the spectra intensity pattern of Dirac cone. **b** Experimental band dispersion along the $\overline{K}$ - $\overline{M}$ - $\overline{K}$ direction. The data in (a) and (b) are probed with LH polarization. **c,d** Same as (a),(b), but measured with the LV polarization. **e** Simulation of the constant energy map above (i) and below (ii) the Dirac cone for the LH polarization, based on the sublattice interference of kagome initial states. **e** Same as (f), but for the LV polarization. The red dashed lines in (a,c) and (e,f) show the Brillouin zone.